\documentstyle[12pt,twoside]{article}
\def\nv{{\bf n}}
\def\rv{{\bf r}}
\def\xv{{\bf x}}
\def\Rv{{\bf R}}
\def\uv{{\bf u}}
\def\ev{{\bf e}}
\def\tauv{{\mbox{\boldmath{$\tau$}}}}
\def\gradv{{\mbox{\boldmath{$\nabla$}}}}
\input epsf
\newcommand{\case}[2]{\mbox{${{\footnotesize #1}\over{\footnotesize #2}}$}}
\def\gradv{\nabla}
\begin{document}
$\qquad$\\
\vspace{0.3in}
\begin{center}
{\bf CHIRALITY IN LIQUID CRYSTALS: FROM MICROSCOPIC ORIGINS TO MACROSCOPIC
STRUCTURE}\\
\vspace{0.3in}
T.C. LUBENSKY, A.B. HARRIS, RANDALL D. KAMIEN, \\
AND GU YAN\\
Department of Physics and Astronomy, University of Pennsylvania,
Philadelphia, PA, 19104
\vspace{0.3in}
\end{center}
\begin{flushleft}
Molecular chirality leads to a wonderful variety of equilibrium structures,
from the simple cholesteric phase to the twist-grain-boundary phases, and
it is responsible for interesting and technologically important materials
like ferroelectric liquid crystals.  This paper will review some recent
advances in our understanding of the connection between the chiral geometry
of individual molecules and the important phenomenological parameters that
determine macroscopic chiral structure.  It will then consider chiral
structure in columnar systems and propose a new equilibrium phase
consisting of a regular lattice of twisted ropes.\\
\vspace{0.2in}
\end{flushleft}
\section{Introduction}
\par
Chirality leads to a marvelous variety of liquid crystalline phases,
including the cholesteric, blue, TGB, and Sm$C^*$ phases, with
characteristic length scales in the ``mesoscopic" range from a fraction of
a micron to tens of microns or more.  In spite of its importance, remarkably
little is known about the connection between chirality at the molecular
level and the macroscopic chiral structure of liquid crystalline phases.
In particular, it is not possible at this time to predict the pitch $P$ (of
order a micron) of a cholesteric phase from the structure of its
constituent molecules.  This talk will address some aspects of the molecular
origins of chirality.  It will also speculate about a possible new phase
in chiral columnar or polymeric systems.
\newpage
\section{What is Chirality?}
\par
\begin{figure}
\centerline{\epsfbox{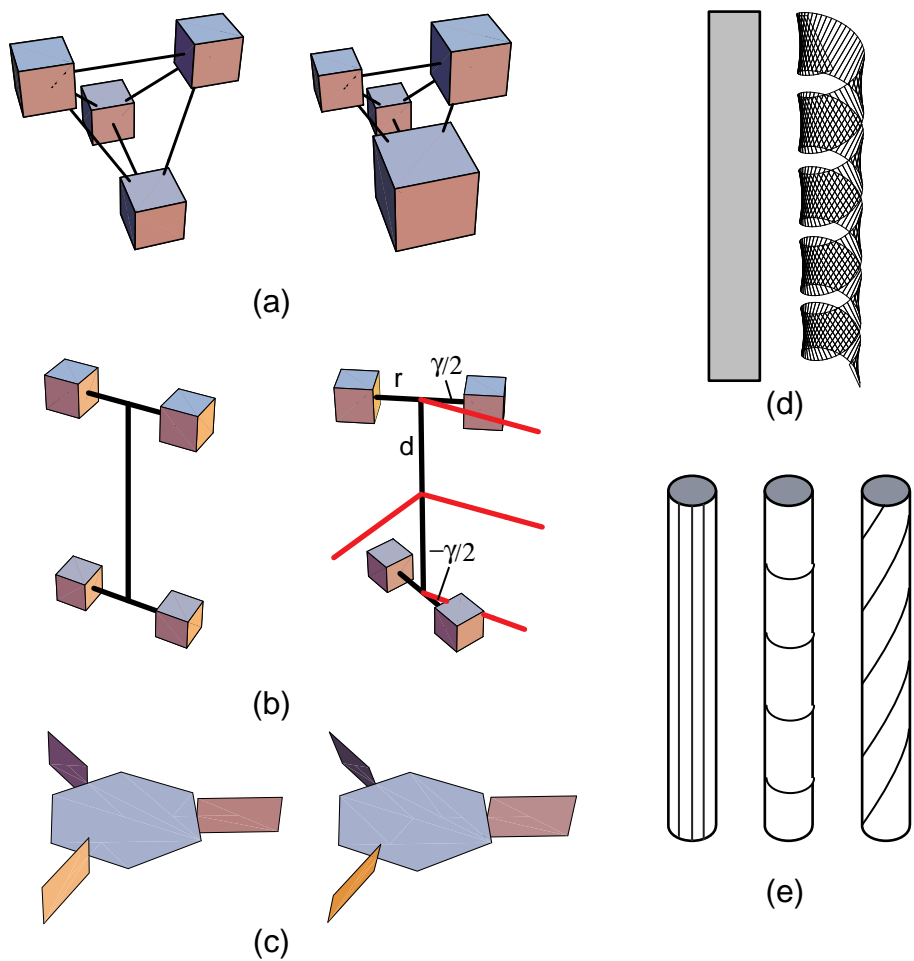}}
\caption{Examples of chiral structures created from achiral ones: (a) left,
achiral tetrahedron with four equal masses at its vertices; right, a chiral
tetrahedron with four unequal masses at its vertices. (b) left, an achiral
planar ``H"; right, a chiral ``twisted H". (c) left, an achiral propeller
with all blades perpendicular to the hexagonal core; right, a chiral
propeller with all blades rotated away from the normal to the hexagonal
plane. (d) left, an achiral planar sheet; right, a helix formed by twisting
a sheet about an cylinder. (e) left and middle, cylinder with achiral
deorations; right, chiral cylinder with helical decorations.}
\label{fig1}
\end{figure}

A molecule is chiral if it cannot be brought into coincidence with
its mirror image\cite{kelvin}.
Examples of chiral molecules and the achiral configurations from which they
are derived are shown in Fig.\ \ref{fig1}.  A chiral molecule must have a
three dimensional structure.  A linear, ``one-dimensional," or a flat,
``planar," molecule cannot be chiral.  In particular a chiral molecule
cannot be uniaxial.  Nevertheless, the local structure of liquid crystals
phases (including the cholesteric and other chiral phases) is nearly
uniaxial.  Indeed the phenomenology of most chiral phases can be explained
in terms of the chiral Frank free-energy density,
\begin{equation}
f = \case{1}{2}K_1 (\gradv \cdot \nv)^2 + \case{1}{2} K_2 [\nv \cdot
(\gradv \times \nv)]^2 + \case{1}{2}K_3 [\nv \times (\gradv \times \nv)]^2
- h \nv \cdot (\gradv \times \nv ) ,
\label{frank}
\end{equation}
expressed in terms of the Frank director $\nv ( \xv )$ specifying the
local direction of uniaxial molecular alignment at the space point $\xv$.
This free energy does not distinguish directly between truly uniaxial
molecules and biaxial molecules that spin about some local axis so that their
average configurations are uniaxial. Chirality in this free energy is reflected
in the term $- h\nv\cdot (\gradv \times \nv )$.  The parameter $h$ is a chiral
or pseudoscalar field that is nonzero only if the constituent molecules
are chiral. It is a phenomenological parameter that may vary with temperature
(or other external field such as pressure) and may even change sign.  Note
that the Frank free energy makes no explicit reference to the fact that a
chiral molecule cannot be uniaxial.  $h$ determines the length scale of
equilibrium chiral structures.  For example, the pitch wavenumber $k_0 =
2 \pi /P$ of a cholesteric is simply $k_0 = h/ K_2$.  Thus the chiral
parameter $h$ can be identified with $K_2 k_0$.  It has units of
energy/(length)$^2$, and one would naively expect it to have a magnitude of
order $T_{NI}/a^2$, where $T_{NI}$ is the isotropic-to-nematic transition
temperature and $a \approx 30$\AA\ is a molecular length.  The elastic
constant $K_2$ does conform to naive expectations from dimensional analysis
and is of order
$T_{NI}/a$. Thus $h = (2 \pi K_2/P) \approx 2 \pi T_{NI}/(aP)$
is or order $[T_{NI}/a^2](a/P)$, which is
$a/P \approx 10^{-3}$ times smaller than naive dimensional
analysis would predict.
\par
The Frank free energy of Eq.\ (\ref{frank}) can be used to treat the
cholesteric phase, chiral smectic phases (when a smectic free energy is
added) and even low-chirality blue phases.
To describe transitions from the isotropic
phase to the cholesteric or blue phases, the Landau-de-Gennes free energy
expressed in terms of the symmetric-traceless nematic order parameter
$Q^{ij}$ is more useful.  When there are chiral molecules, this free
energy is
\begin{eqnarray}
f_Q & = & \case{1}{2} r {\rm Tr} Q^2 - w {\rm Tr} Q^3 + u {\rm Tr} Q^4
\nonumber\\
& & +\case{1}{2} C_1 \partial_i Q^{jk} \partial_i Q^{jk}
+\case{1}{2} C_2 \partial_i Q^{ik} \partial_j Q^{jk}
+ H \epsilon^{ijk}Q^{il}\partial_j Q^{kl} .
\label{LdG}
\end{eqnarray}
The chiral term in this expression is the one proportional to the Levi-Civita
symbol $\epsilon^{ijk}$.  Indeed, the usual mathematical signal of chiral
symmetry breaking in a free energy is a term linear in $\epsilon^{ijk}$.
It is present in the Frank free energy in the term $\nv \cdot (\gradv
\times \nv) = \epsilon^{ijk}n^i\partial_j n^k$.  In the nematic phase
the order parameter $Q_{ij}$ is unixial:
$Q^{ij} = S ( n^i n^j - \case{1}{3}\delta^{ij})$, where $S$ is the
Maier-Saupe order parameter, and Eq.\ (\ref{LdG}) reduces to
Eq.\ (\ref{frank}) with $h\sim H S^2$.  Thus, like the Frank free
energy, the Landau-de-Gennes free energy makes no explicitly reference to
biaxiality.
\par
A theoretical goal should be the calculation of the parameter $h$
(or $H$) from
molecular parameters and inter-molecular interactions.  A theory should (1)
show why $h$ is smaller than naive arguments would predict, (2) provide at
least semi-quantitative guidance as to how molecular architecture affects
$h$, and (3) elucidate the transition from necessarily non-uniaxial chiral
molecules to an essentially uniaxial macroscopic free energy.  Since $h$ is
zero for achiral molecules, it is natural to expect that $h$ will be
proportional to some parameter measuring the ``degree of chirality" of
constituent molecules.  Section III will introduce various measures of
molecular chirality and discuss why it is impossible to define a single
measure of chirality.  $h$ should depend on interactions between molecules
as well as on chiral strength.  Section IV will present a calculation of
$h$ for simple model molecules composed of atoms interacting via classical
central force potentials\cite{HarKam97}.
This calculation will highlight the role of biaxial
correlations.  It ignores, however, chiral quantum mechanical dispersion
forces\cite{Vander76}.  Finally, Sec. V will discuss chirality
in columnar systems and
speculate on the possibility of a new chiral phase consisting of a lattice
of twisted ropes.
\section{Measures of Molecular Chirality}
\par
To keep our discussion as simple as possible, we will consider only
molecules composed of neutral atoms.  All information about the symmetry
of such a molecule can be constructed from its mass density
relative to its center of mass:
\begin{equation}
\rho ( \xv ) = \sum_{\alpha} m_{\alpha} \delta ( \xv - \rv_{\alpha} ),
\end{equation}
where $m_{\alpha}$ is the mass of atom $\alpha$ whose position vector
relative to the molecular center of mass is $\rv_{\alpha}$.  A molecule is
chiral if there exists no mirror operation $M$ under which $\rho(\xv)$ is
invariant, i.e., for all $M$, $\rho(\xv) \neq \rho ( M \xv )$.
Mass moment tensors contain information about the symmetry
and mass distribution of a molecule, and they can be used to construct
chiral measures.  The first moment specifies the position of the center of
mass, which is uninteresting since we can always place the center of mass
at the origin of our coordinates.  Second- and third-rank moments are,
however, interesting, and we define
\begin{eqnarray}
C^{ij}& = &\sum_{\alpha} m_{\alpha} \left( r_{\alpha}^i r_{\alpha}^j -
\case{1}{3}r_{\alpha}^2 \delta^{ij} \right) , \nonumber\\
C_n^{ij}& = &\sum_{\alpha} m_{\alpha} (r_{\alpha})^n\left( r_{\alpha}^i
r_{\alpha}^j -\case{1}{3}r_{\alpha}^2 \delta^{ij} \right) ,
\end{eqnarray}
and
\begin{eqnarray}
D^{klm} & = & \sum_{\alpha} m_{\alpha}\left[r_{\alpha}^k r_{\alpha}^k
r_{\alpha}^l - \case{1}{5} (r_{\alpha})^2  (r_{\alpha}^k \delta^{lm} +
r_{\alpha}^l \delta^{km} + r_{\alpha}^m \delta^{kl} ) \right] , \nonumber \\
D_n^{klm} & = & \sum{\alpha} m_{\alpha}(r_{\alpha})^n \left[r_{\alpha}^k
r_{\alpha}^k r_{\alpha}^l - \case{1}{5} (r_{\alpha})^2  (r_{\alpha}^k
\delta^{lm} + r_{\alpha}^l \delta^{km} + r_{\alpha}^m \delta^{kl} ) \right]
{}.
\end{eqnarray}
where $(r_{\alpha})^n$ is $(\rv_{\alpha} \cdot \rv_{\alpha})^{n/2}$.
These tensors are symmetric and traceless, and they are invariant under
the mirror operation that interchanges any two axes, say $x$ and $y$.
Therefore, they cannot by themselves encode any information about molecular
chirality.
\par
To construct a quantity that is sensitive to chirality, we need
combinations involving three distinct tensors in combination with the
anti-symmetric Levi-Civita symbol $\epsilon^{ijk}$.  One possible chiral
parameter is \\
$\epsilon^{ijk} D^{ilm} C^{jl} (C^2 )^{km}$, where the
Einstein convention on repeated indices is understood, and where
$(C^2)^{km}$ is the $km$ component of the tensor $C^2$.
More useful chiral measures can be constructed by dividing $C^{ij}$
into a uniaxial and a biaxial part.  The tensor $C^{ij}$ has five
independent components that can be parametrized by two eigenvalues
$\psi_Q$ and $\psi_B$ and an orthonormal triad $\{\ev_1, \ev_2, \ev_3
\}$ specifying the directions of the principal axes of the molecule.
Thus, we can write
\begin{equation}
C^{ij} = Q^{ij} + B^{ij} ,
\end{equation}
where
\begin{equation}
Q^{ij} = \psi_Q (e_3^i e_3^j - \case{1}{3} \delta^{ij} ) \equiv \psi_Q
{\tilde Q}^{ij}
\label{tildeQ}
\end{equation}
and
\begin{equation}
B^{ij} = \psi_B (e_1^i e_1^j - e_2^i e_2^j ) \equiv \psi_B {\tilde
B}^{ij} .
\label{tildeB}
\end{equation}
The last equation defines the reduced biaxial tensor ${\tilde B}^{ij}$ that
depends only on the principal axis vectors $\ev_1$ and $\ev_2$.
For a uniaxial molecule, the parameter $\psi_B$ is zero, and we will
refer to $Q^{ij}$ as the uniaxial part of $C^{ij}$ and $B^{ij}$ as the biaxial
part (even though the equilibrium average $\langle Q^{ij} \rangle$ can develop
a biaxial part). The third rank tensor $D^{ijk}$ has seven independent
components, which can be represented in the basis $\{\ev_1 , \ev_2 , \ev_3 \}$
defined by the principal axes of $C^{ij}$.  Of these, only one component is of
interest to the present discussion:
\begin{equation}
D^{klm} = \psi_C ( e_1^k e_2^l e_3^m + 5 \,\,{\rm perm} ) \equiv \psi_C
{\tilde D}^{klm} .
\label{tildeD}
\end{equation}
This equation defines the reduced third-rank tensor ${\tilde D}^{ijk}$.
We can now introduce the chiral strength\cite{HarKam97}
\begin{equation}
\psi = \epsilon^{ijk} e_3^i e_3^l B^{jm} D^{klm} = 2 \psi_B \psi_D =
D^{xyz} (B^{xx} - B^{yy} ) .
\label{chiralstr}
\end{equation}
This parameter clearly changes sign under any mirror operation.  It must,
therefore, be zero for any achiral molecule.  It is a continuous function
that has opposite signs for structures of opposite chirality.  Consider,
for example, the following mirror operations: (1) $z\rightarrow -z$,
$D^{xyz}$ changes signs but $B^{xx} - B^{yy}$ does not; (2) $x
\leftrightarrow y$, $(B^{xx} - B^{yy})$ changes sign but $D^{xyz}$ does not.
More generally, we could introduce a class of chiral strength
parameters defined in
terms of $D_n^{klm}$ and the biaxial part, $B_n^{ij}$, of the tensor
$C_n^{ij}$:
\begin{equation}
\psi_{pn} = \epsilon^{ijk} e_3^i e_3^l B_p^{jm} D_n^{klm} .
\end{equation}
The parameter $\psi_{00}$ is equal to $\psi$.  For all other $p$ and $n$,
$\psi_{pn}$ is distinct from $\psi$.
\par
It is instructive to calculate $\psi$ for some of the chiral molecules
shown in Fig.\ \ref{fig1}.  The ``twisted H" structure is the simplest
example of a geometrically chiral object.  It has the pedagogical virtue
that its geometric properties depend in a simple continuous way on the
angle $\gamma$ between arms.  When $\gamma = 0$ or $\gamma = \pi/2$, the H
is achiral.  Thus any chiral measure must be zero at these points,
and our expectation is that chiral measures will go continuously to zero at
these points.  Let
$r$ be the radius of the arms of the ``H" and $d$ be half its height.  Then
$\psi_Q = 2 ( 2d^2 -r^2)$,
$\psi_B = 2 r^2 \cos \gamma$, $\psi_D = 2 d r^2 \sin\gamma$, and
\begin{equation}
\psi = 4 d r^4 \sin 2 \gamma .
\end{equation}
As required, $\psi\rightarrow 0$ as $\gamma \rightarrow 0, \pi/2$.
Another interesting example is the helix or spiral structure in Fig.\
\ref{fig1}(d).  Parametrizing the helix by its position as a function of
normalized arclength $s$: $\Rv ( s) = [r \cos ( 2 \pi ns ) , r \sin ( 2 \pi
n s ), Ls]$, where $n$ is an integer, we find
\begin{equation}
\psi \approx {3 r^4 L\over (2 \pi n )^3} \left[1 - {24 \over ( \pi n)^2}
\right] .
\end{equation}
Thus, $\psi$ decreases as the helix becomes more tightly wound
($n$ increases) - a reasonable result because at large distances a
tightly wound spiral looks more like a uniform cylinder than does a not so
tightly wound one.
\par
There are many, in fact infinitely many, chiral strength parameters that
provide a measure of the degree of chirality of a molecule.  The set
$\psi_{pn}$ defined above is just one of infinitely many sets we can
define.  At first, it may seem strange that the degree of chirality may be
characterized in so many different ways.  After all, chirality is simply
the absence of a mirror symmetry, which involves a discrete operation.
We might, therefore, have expected a single ``chiral order parameter"
analagous to the order parameter for an Ising model, which also describes
the absence of a discrete symmetry.  Chirality is, however, a function of
the positions of all of the atoms in a molecule, which cannot be
characterized by a single variable.  Thus, chirality is perhaps more like
the absence of spherical symmetry.  There an infinite number of parameters,
the spherical harmonics for example, that chracterize deviations from
perfect sphericity.
\par
Some measures of chirality are more useful than others,  As we will see, the
large-distance potential between two generic chiral molecules is directly
proportional to the strength $\psi$ introduced in Eq.\ (\ref{chiralstr}).
Thus, $\psi$ provides a good measure of the strength of chiral interactions
and enters into the calculations of $h$.  There are molecules, such as the
chiral tetrahedron of Fig.\ \ref{fig1}a, for which $\psi$ is zero.  For
such molecules, $\psi$ is clearly not very useful.  Other measures are,
however, nonzero and provide a measure of the strength of chiral
interactions.
\section{Chiral Pathways between Enantiomers}
\par
The mirror image of a chiral molecule is called its enantiomer.  Any
measure of chirality will change sign under a mirror operation, and as a
result, chiral enantiomers will have chiral strengths of equal magnitude but
opposite sign.  This property implies that any series of continuous distortions
of a molecule that takes it from a chiral configuration to that of its
enantiomer must necessarily pass through a point where any given chiral
measure is zero.  It is tempting to conclude that this point corresponds to
an achiral configuration and that any pathway between chiral enantiomers
must pass through an achiral state.  This conclusion is false.  There are
continuous pathways between chiral enantiomers all of whose states are
chiral (for further discussion of this and the concept of a ``topological
rubber glove," see \cite{Walba95}.
A molecule is achiral if and only if any and all chiral measures
are zero.  For a particular chiral configuration, a given chiral
measure may be zero, but at least one chiral measure be must be nonzero.
\begin{figure}
\centerline{\epsfbox{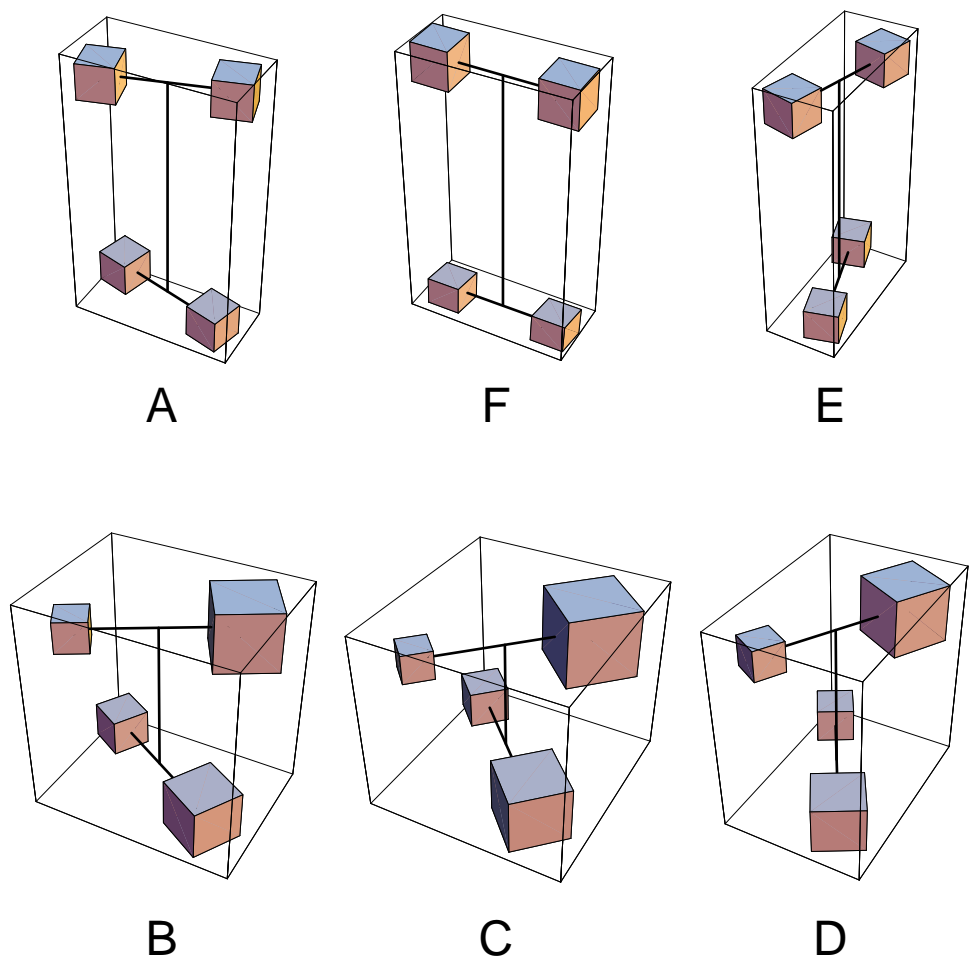}}
\caption{Pathways between chiral enantiomers consisting of $4$ atoms at the
vertices of an H.  $A$ and $E$ are chiral enantiomers (mirror images).  The
$4$ masses and their positions can be changed continuously to provide
continuous pathways between the two enantiomers.  Path $AFE$ passes through
the achiral planar configuration $F$.  Path $ABCDE$ passes through chiral
configurations only.  Configuration $C$ is the chiral tetrahedron with
unequal masses.}
\label{fig2}
\end{figure}
\par
\begin{figure}
\centerline{\epsfbox{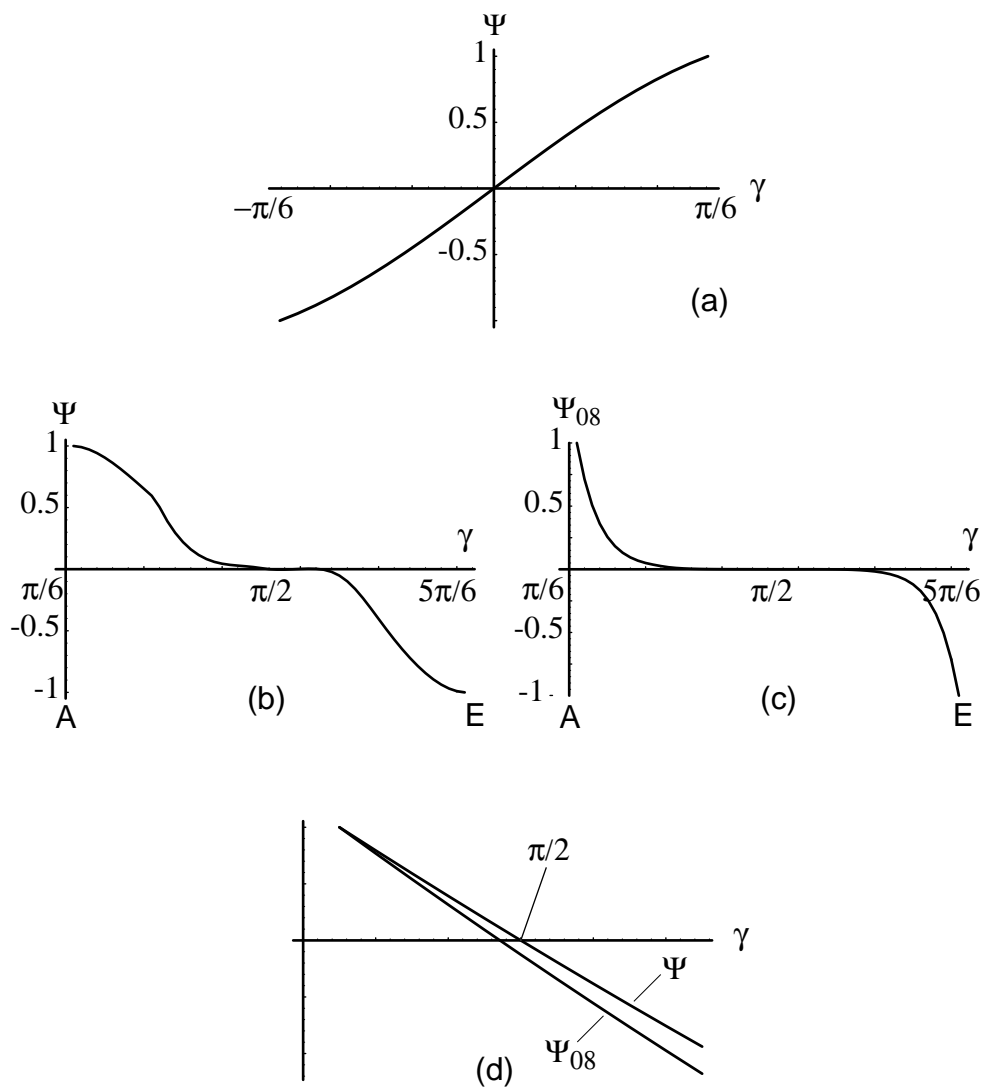}}
\caption{Plots of reduced chiral strength parameters for different pathways
between chiral enantiomers shown in Fig.\ \protect{\ref{fig2}}. (a) $\Psi =
\psi_{00}/\psi_{00} (A)=\psi_{08}/\psi_{08} (A)$ for the path $EFA$ passing
through the achiral configuration $F$.  (b) $\Psi = \psi_{00}/\psi_{00}(A)$
for the path $ABCDE$. (c) $\Psi_{08} = \psi_{08}/\psi_{08}(A)$ for the path
$ABCDE$. (d) $\Psi$ and $\Psi_{08}$ in the vicinity of $\gamma = \pi/2$.
They pass through zero at different points.}
\label{fig3}
\end{figure}

To illustrate this point, let us consider the pathways depicted in Fig.\
\ref{fig2} between reflected images of a twisted H.  $A$ is a twisted H
with an angle $\gamma = \pi/6$ between top and bottom arms.  $E$ is its
chiral enantiomer with angle $- \pi/6$ between arms. $F$ is the the
achiral planar structure.  $B$ through $D$ are all chiral configurations with
different masses at the vertices of the H.  $C$ is the chiral tetrahedron
with four unequal masses.  The path $AFE$ between chiral enantiomers
passes through the achiral configuration $F$ for which all chiral measures
are zero.  On the other hand, all configurations in the path ABCDE are
chiral even though $A$ and $E$ are chiral enantiomers for which any chiral
measure $\psi$ must have opposite signs: $\psi (E) = - \psi (A)$.  Thus,
any continuous chiral measure must pass through zero at some point on the
path $ABCDE$, but different chiral measures do not have to pass through
zero at the same point. To illustrate these facts, we consider a specific
parametrization of the closed path $ABCDEFA$.  Let the four masses in the
twisted H be
\begin{equation}
\{m_1,m_2,m_3,m_4\} = m[\{1,1,1,1\} + \mu \{0.31, - 0.31, 0.23, -0.23\}].
\end{equation}
The configurations of the twisted H are thus specified by the three
parameters $\mu$, $d/r$, the ratio of the semiheight to the arm radius,
and $\gamma$, the angle between the arms of the H.  The
following properties of a chiral measure $\psi$ are apparent:
\begin{enumerate}
\item $\mu = 0$: Equal masses
\begin{enumerate}
\item $\gamma = 0$, $\gamma = \pi/2$: achiral $\Rightarrow$ $\psi = 0$
\item $d = r \sin (\gamma/2)$ or $d = r \cos (\gamma /2)$: achiral
$\Rightarrow \psi = 0$
\item $\psi ( \gamma ) = - \psi ( - \gamma ) = - \psi ( \pi - \gamma ) =
2 d r^4 \sin 2 \gamma $.
\end{enumerate}
\item $\mu \neq 0$: Unequal masses
\begin{enumerate}
\item $\gamma = 0$: planar achiral molecule $\Rightarrow \psi = 0$
\item $\gamma \neq 0$: chiral molecule $\Rightarrow \psi \neq 0$
\item $\psi ( \gamma , \mu ) = \psi (\gamma, - \mu )$
\item $d = r / \sqrt{2}$: chiral tetrahedron with unequal masses.
\end{enumerate}
\end{enumerate}
The points in Fig. \ref{fig2} correspond to the following values of
$[\gamma, \mu , d/r]$: $A = [\pi/6,0,1]$, $B = [\pi/3,3/4,(\sqrt{2} +
3)/(4 \sqrt{2})]$, $C = [0,1,1/\sqrt{2}]$, $D = [2 \pi/3, 3/4, (\sqrt{2} +
3)/(4\sqrt{2})]$, $ E = [5 \pi/6, 0, 1 ] = [- \pi/6,0,1]$.
Figure \ref{fig3}a shows
the normalized chiral strengths $\Psi_{nm} = \psi_{nm}(\gamma, \mu ,
d/r)/ \psi(A)$ for $(n,m) = (0,0)$ and $(n,m) = (0,8)$
as a function of $\gamma$ for the path $EFA$ with
$\mu = 0, d/r = 1$.  Only one curve appears in the figure because the
two curves for $\Psi_{00}$ and $\Psi_{08}$ are identical.
As required, the curve passes through zero at
the achiral point $\gamma = 0$.  Figures \ref{fig3}a and
\ref{fig3}b show the normalized
strengths $\Psi = \Psi_{00}$ and $\Psi_{08}$ for the path $ABCDE$ with
$\pi/6 \leq \gamma \leq 5 \pi/6$ with $\mu = 1 - (9/4)[(2 \gamma
/\pi)-1]^2$ and $d/r = (1 \sqrt{2}) + (9/4)[1 - (1/ \sqrt{2})][(2
\gamma/\pi) - 1]^2$.  The two curves are clearly different, but they
appear to go to zero at $\gamma = \pi/2$.  Figure \ref{fig3}d shows a
blowup of the region around $\gamma = \pi/2$, showing that $\Psi$
and $\Psi_{08}$ pass through zero at different values of $\gamma$.  Note
that $\Psi$ passes through zero at $\gamma = \pi/2$ indicating that this
parameter does not provide a measure of the chirality
of the tetrahedron with unequal masses.  $\Psi_{08}$ is nonzero but small
at $\gamma = \pi/2$.
\par
The path $ABCDE$ was chosen to pass through the chiral tetrahedron at point
$C$.  A tetrahedrally coordinated carbon atom bonded to four differenct
atoms or chemical groups is a chiral center.  It is used to identify
chriality in molecules.  Interestingly, the simple chiral tetrahedron is
not very chiral according to the measures $\psi_{pn}$.  All measures
$\psi_{00}$ to $\psi_{05}$ are zero for this molecule.  This is a partial
explanation of why the difference between $\Psi$ and $\Psi_{08}$ is so
small near $\gamma = \pi/2$.  Other pathways between $A$ and $E$ would
display more pronounced differences.
\section{Calculation of $h$}
\par
Two distinct microscopic mechanisms contribute to a non-vanishing macroscopic
chiral parameter $h$.  The first is quantum mechanical in origin and has its
clearest manifestation in the generalization of the familiar Van der Waals
dispersion potential to chiral molecules\cite{Vander76}.
The electrostatic potential
between molecules is expanded to include dipole-quadrupole as well as
dipole-diple interactions.  The expectation value of the
dipole-moment--quadrupole-moment product $\langle p^i Q^{jk} \rangle$ is
nonzero for a chiral molecules and is proportional to $(\epsilon^{ijl} e_3^l
e_3^k + {\rm perm} )$ for a molecule spun about the direction $\ev_3$.  The
resultant potential has the form,
\begin{equation}
U = V( R ) \ev_3 \cdot \ev_3^{\prime} \ev_3 \cdot ( \Rv \times \ev_3^{\prime}
) ,
\end{equation}
where $R$ is the spatial separation of the centers of mass of the two
molecules and $V(R) \sim R^{-7}$.
This potential leads directly to the
Landau-de-Gennes chiral interaction $\epsilon^{ijk}Q^{il}\partial^j Q^{kl}$
in the isotropic phase [Eq.\ (\ref{LdG})] and to the
Frank chiral term in the nematic phase with
$h \sim S^2$.  The calculation of the strength of the potential $V(R)$ for
real liquid crystal molecules is complex.  The second mechanism can be
described in purely classical terms.  It arises from central force potentials,
including hard-core potentials, between points (or atoms) on molecules with
chiral shapes.  This is the mechanism popularized by Straley's\cite{Straley}
image of two interlocking screws that twist relative to each other.  Here we
will consider only the second mechanism. We will show that $h$ is nonzero
only if there are biaxial correlations between molecules and as a consequence
is zero in any mean-field theory with uniaxial order parameters.
\par
We consider each molecule $A$ to be composed of atoms $\alpha$ whose
positions within the molecule are rigidly fixed.  Atoms on different
molecules interact via a central potential $V(R)$.
Let $\Rv_A$ be the position of the center of mass of molecule $A$,
$\rv_{A\alpha}$ be the position of atom $\alpha$ relative to $\Rv_A$, and
$\Rv_{A\alpha} = \Rv_A + \rv_{A\alpha}$ be the position of atom $\alpha$ in
molecule $A$.
The total intermolecular potential energy is thus
\begin{equation}
U=\case{1}{2} \sum_{A\alpha \neq B\beta} V(|\Rv_{A\alpha} - \Rv_{B\beta} |).
\end{equation}
The chiral parameter $h$ is the derivative of the free energy density with
respect to the cholesteric wave number $k = 2 \pi /P$ evaluated in the
aligned state with spatially uniform director $\nv$ and $k=0$:
\begin{equation}
h = - \left.{\partial f \over \partial k}\right|_{k=0}=
-{1 \over \Omega}\left.\left\langle {\partial U\over \partial k}
\right\rangle\right|_{k=0}=-{1 \over 4 \Omega}\sum_{BA}T_{BA} ,
\label{hh}
\end{equation}
where $\Omega$ is the system volume and
\begin{equation}
T_{BA} = \left\langle \sum_{\beta\alpha} \epsilon^{ijk} R_{\perp}^i
r_{\beta}^j\partial^k V(\Rv + \rv_{\beta} - \rv_{\alpha} )\right\rangle
\end{equation}
is the ``projected torque", $\Rv_{\perp} \cdot \tauv_{AB}$,
where $\tauv_{AB}$ is
the torque exerted on $A$ by $B$. Here $\rv = \Rv_B - \Rv_A$ and
$\rv_{\perp}$ is the component of $\rv $ perpendicular to the uniform
director $\nv$.
This expression for $h$ is quite
general.  It can be used in Monte Carlo and molecular dynamics simulations.
It is valid both classically and quantum mechanically. (It will reproduce
the dispersion calculations described above if electron as well as atomic
coordinates are included).  Here we restrict our attention to classical
central forces.  The fundamental question is, how do nonchiral central forces
give rise to effective chiral forces?  To answer this question, we consider
the multipole expansion of $U$.  Schematically, we have
\begin{eqnarray}
\sum_{\alpha, \beta} V ( \Rv_{A\alpha} - \Rv_{B\beta} ) & \rightarrow &
\case{1}{4}[\partial^i \partial^j \partial^k \partial^l V] C_A^{ij} C_B^{kl}
\\
& & + \case{1}{24}[\partial^i\partial^j\partial^k\partial^l\partial^m
V] (C_A^{ij}D_B^{klm} + C_B^{ij} D_A^{klm}) . \nonumber
\label{mm}
\end{eqnarray}
The second term on the right hand side of this equation is the chiral term.
It can be expressed as
\begin{equation}
V_C = \case{1}{24}\psi_D \psi_B [\partial^i\partial^j\partial^k\partial^l
\partial^m V] ( {\tilde B}_A^{ij} {\tilde D}_B^{klm} +
{\tilde B}_B^{ij} {\tilde D}_A^{klm} ) ,
\end{equation}
where the reduced tensors ${\tilde B}^{ij}$ and ${\tilde D}^{klm}$
are defined in Eqs.\ (\ref{tildeB}) and (\ref{tildeD}).
The triad of vectors $\{\ev_1,\ev_2,\ev_3\}$ forms a right-handed coordinate
system.  Since this triad is determined by the second-moment tensor, it can
be chosen to be invariant under a chiral operation $M$.  Thus, the reduced
tensors ${\tilde B}^{ij}$ and ${\tilde D}^{klm}$ do not change sign under a
chiral operation.  The chiral parameter $\psi = 2 \psi_B \psi_D$ and the
potential $\partial^5 V$ both change sign under a chiral operation on all
atoms.  Thus the potential $V_C$ is invariant under a chiral operation
involving {\em all} atoms.  If, however, we perform a chiral operation
about the center of mass of each molecule (i.e., convert each molecule to
its enantiomer) leaving the centers-of-mass
positions fixed, $\psi$ changes sign but $\partial^5 V$ remains unchanged.
Thus, the potential $V_C$ changes sign upon reversing the chirality of
constituent molecules.  As we noted
earlier, any term linear in the Levi-Civita symbol $\epsilon^{ijk}$ is
chiral.  The potential $V_C$ is linear in $\epsilon^{ijk}$ because the triad
$\{\ev_1, \ev_2, \ev_3\}$ is right handed, and $\ev_1 = (\ev_2 \times
\ev_3)$, $\ev_2 = (\ev_3 \times \ev_1)$, and $\ev_3 = (\ev_1\times \ev_2)$.
Thus
\begin{equation}
{\tilde D}^{ijk}= \epsilon^{ilm}e_3^j e_3^l {\tilde B}^{km}
+\epsilon^{klm} e_3^i e_3^l {\tilde B}^{jm} + \epsilon^{jlm}e_3^k e_3^l
{\tilde B}^{im} .
\end{equation}
\par
We now consider the evaluation of $h$ from Eq.\ (\ref{hh}) and the multipole
expansion of Eq.\ (\ref{mm}).  To simplify our calculation, we restrict all
molecular axes $\ev_{3A}$ to be rigidly aligned along a common axis $\nv_0$.
This is equivalent to forcing the Maier-Saupe order parameter $S$ to be
unity.  In this case, the potential $V_C$ involves only the product of the
biaxial moment ${\tilde B}^{ij}$ on different molecules, and upon averaging,
the torque becomes
\begin{equation}
T_{BA} = \psi \langle K(R) \Gamma_B ( R) \rangle ,
\end{equation}
where
\begin{equation}
\Gamma_B ( R ) = \langle {\tilde B}_B^{ij} {\tilde B}_A^{ij}
\rangle = 2 \langle \cos 2 ( \phi_B - \phi_A ) \rangle
\end{equation}
is the biaxial correlation function and
\begin{equation}
K(R) = R_{\perp}^2 \left\{ g^{(3)} + R_{||}^2 g^{(4)} + \case{1}{4}
R_{\perp}^2 [g^{(4)} + R_{||}^2 g^{(5)}]\right\} ,
\end{equation}
where $g(R^2/2)= V(R)$, and $g^{(n)}(x) = d^n g (x) /dx^n$.
If molecules rotate independently, the biaxial correlation function
$\Gamma_B(\Rv)$ is zero.  Mean-field theory seeks the best single particle
density matrix and thereby ignores correlations between different particle.
Thus, in mean-field theory, $\Gamma_B$ is zero, and the chiral potential
$h$ is zero unless there is long-range biaxial order with a nonvanishing
value of $\langle \cos 2 \phi_A \rangle$. The latter observation
was made some time ago\cite{Schroder,Goossens}. Nevertheless, there have
since been mean-field calculations claiming to produce a nonvanishing $h$
in a uniaxial system\cite{evans}.
In uniaxial systems, $\Gamma_B ( R) \propto e^{-R/\xi}$
where $\xi$ is the biaxial correlation, which should be order a molecular
spacing.  We can now provide a partial answer to the question of why $h$ is
so much smaller than naive scaling arguments would predict.  First, the
chiral strength $\psi$ may be quite small, particularly for tightly would
helical molecules such as DNA or tobacco mosaic viruses.  Second, the
biaxial correlation length could be small (because, for example the
molecules are only weakly biaxial), allowing only very near neighbors
to contribute to the sum over $AB$ in Eq.\ (\ref{hh}) for $h$.
\section{Chiral Columnar Phases}
\par
\begin{figure}
\centerline{\epsfbox{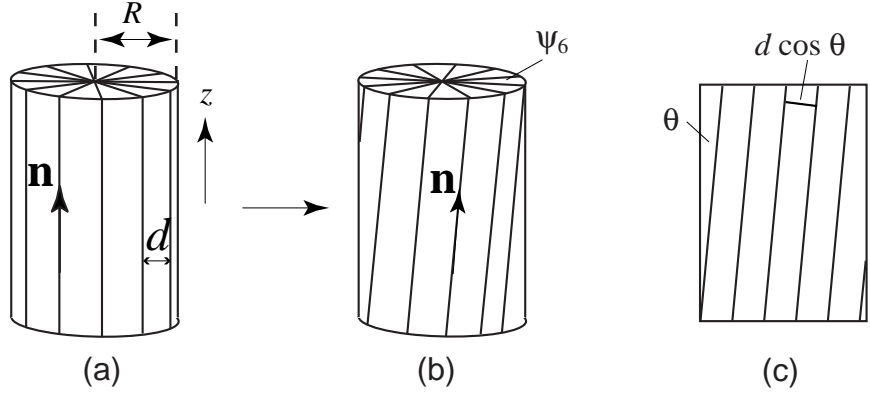}}
\caption{Twisting a cylinder of a columnar lattice.  (a) The untwisted
cylinder of radius $R$ with equilibrium spacing between columns $d$. (b) The
twisted cylinder. (c) The outer surface of the twisted cylinder unrolled.  The
columns make an angle $\theta =\tan^{-1} k_0 R$ with respect to the
cylindrical axis.  The normal spacing between between layers is now
$d/\cos \theta \approx d(1 + \theta^2/2 )$.}
\label{fig4}
\end{figure}
\begin{figure}
\centerline{\epsfbox{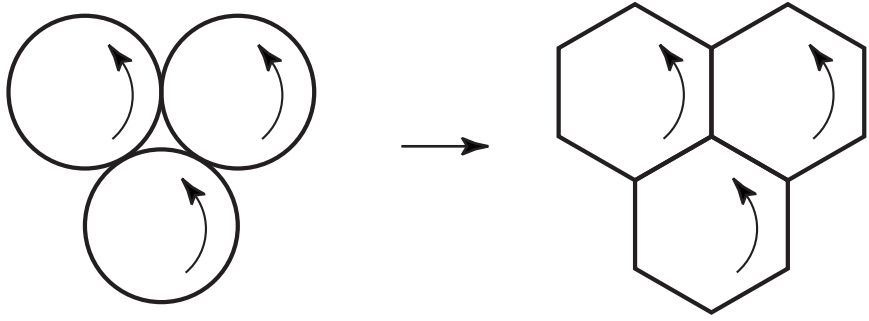}}
\caption{Close-packed twisted cylinders deformed to spacefilling hexagonal
twisted cylinders.}
\label{fig5}
\end{figure}

We now have a fairly complete catalog and understanding of equilibrium phases
of rod-like chiral mesogens.  Our understanding of chiral phases of disc-like
chiral mesogens is less complete.  Such mesogens have been
synthesized\cite{MalJac82,BokHel95a}.
They produce cholesteric and columnar mesophases, the latter of which
exhibit unusual spiral textures\cite{MalJac82}.
Columnar phases with spontaneous chiral symmetry breaking in which
propeller-like molecules rotate along columns have been
observed\cite{FonHei88}. In addition, long, semi-flexible polymers
like DNA whose natural low-temperature phase has columnar
symmetry, exhibit cholesteric, columnar, and blue phases\cite{Livolant91},
and even what appears to be a hexatic phase\cite{Pod96}.  There
have been theoretical predictions of a number of phases including (1) a
moir\'{e} phase\cite{KamNel95} in which the hexagonal columnar
lattice rotates in discrete
jumps about an axis parallel to the columns across twist grain boundaries
consisting of a honeycomb lattice of screw dislocations, (2) a tilt-grain
boundary phase\cite{KamNel95} in which columns
rotate in discrete jumps bout an axis perpendicular to the columns
across twist-grain
boundaries consisting of a parallel array of screw dislocations, and (3) a
soliton phase\cite{YanLub97} in which molecules all rotate in the same sense in
their
columns without disrupting the regular hexagonal columnar lattice.  Many
other states are possible. Here we will consider one of them: a hexagonal
lattice of twisted ropes.  This phase would be constructed
as follows:  First, take cylindrical sections of radius $R$ of a columnar
lattice and twist them about an axis parallel to the columns as shown in
Fig.\ \ref{fig4}.  This operation will cost strain energy but will gain twist
energy.   Next put these columns on a regular hexagonal lattice and deform the
cylinders to a hexagonal shape to produce a dense packing as shown in Fig.
\ref{fig5}.  The interface between adjacent hexagonal cylinders
is a twist grain boundary
and is favored if chiral interactions are sufficiently strong.  On the other
hand, there is a strain energy cost associated with distorting the cylinders
and an energy cost associated with the vertices of the hexagons.  The radius
of the original cylinders (and thus the lattice spacing) and the degree of
twist within a cylinder will adjust to minimize the free energy.  This lattice
of twisted ropes should be competitive with the moir\'{e}, tilt-grain-boundary,
and soliton phases.
\par
To be more quantitative, we now investigate in more detail the various
contributions to the energy of this lattice.  The elastic energy density
of a columnar phase with columns aligned along the $z$ axis is
\begin{equation}
f_{\rm el} = \case{1}{2} \lambda u_{\alpha\beta}^2 + \mu u_{\alpha\beta}
u_{\alpha\beta} + \case{1}{2} K_3 (\partial_x \uv )^2 + \case{1}{2} K_6
(\gradv \theta_6 )^2 ,
\label{elastenergy}
\end{equation}
where
\begin{equation}
u_{\alpha \beta} = \case{1}{2} (\partial_{\alpha} u_{\beta}
\partial_{\beta} u_{\alpha} + \partial_{\gamma} u_{\alpha} \partial_{\gamma}
u_{\beta} - \partial_z u_{\alpha} \partial_z u_{\beta}) , \,\, \alpha, \beta
= x, y
\end{equation}
is the nonlinear strain for a columnar lattice in mixed Euler-Lagrangian
coordinates.  The first three terms in this nonlinear
strain are identical to the nonlinear Lagrangian strain (with a relative plus
sign between the linear and nonlinear terms) of a two-dimensional
solid in which the coordinates $x$ and $y$ refer to positions in a reference
unstrained sample.  The third term reflects the three-dimensional nature of the
columnar phase.  There is a minus sign between it and the preceding terms
because $z$ measures a coordinate in space and not a coordinate in a
reference solids;  it is an Eulerian rather than a Lagrangian
variable\cite{ChaLub95}.
The variable $\theta_6 = (\partial_x u_y - \partial_y u_x)/2$ in Eq.\
(\ref{elastenergy}) specifies the direction of hexatic order of the lattice.
In a chiral system, there are additional chiral terms in the free energy
density:
\begin{equation}
f_{\rm ch} = - \gamma_{\theta} \partial_z \theta_6 - \gamma_n \nv \cdot
( \gradv \times \nv ) ,
\end{equation}
where $\nv$ is the nematic director, which is parallel to the local columnar
axis.  The first term in $f_{\rm ch}$ favors twisting of the hexagonal
columnar lattice along the columns, i.e.,
it favors a twisted rope or the moir\'{e} configuration.
The second term favors twisting of the columns along directions perpendicular
to the local columnar axes, i.e., it favors a tilt-grain-boundary structure.
\par
We can use $f_{\rm el}$ and $f_{\rm ch}$ to calculate the preferred
configuration of a cylinder of a chiral columnar lattice of radius $R$ in
which dislocations and other defects are not allowed.  The elastic energy of
a lattice in which $\theta_6 = k_0 z$ is uniformly twisted with twist
wavenumber $k_0$ is controlled by the nonlinear term $\partial_z u_{\alpha}
\partial_z u_{\beta}$ of the strain.  Indeed the two-dimensional Lagrangian
part of the strain arising from such a uniform twist is zero! There are
nonlinear contributions to the strain, however, because the distance between
neighboring columns increases as the columns are twisted as depicted in Fig.\
\ref{fig4}.  Since twisting produces no linear strain, the energy required to
twist a columnar lattice is lower than that required to twist a
three-dimensional elastic solid.  The elastic energy per unit length of a
twisted columnar lattice can be calculated.  For small $k_0 R$, it is
\begin{equation}
{E_{\rm el}\over L} = {\overline\mu} (k_0 R)^4 \pi R^2 + \case{1}{4} K_3
k_0^2 (k_0 R)^2 \pi R^2 + \case{1}{2}K_6 k_0 ^2 \pi R^2 .
\label{bulk}
\end{equation}
where $\overline\mu$ is an effective modulus that is strictly proportional to
$\mu$ in an incompressible system where the bulk coefficeint $\lambda$ is
infinite. This energy should be compared to the result,
$\mu k_0^4 R^4$, for a three-dimensional solid. The chiral
energy per unit length is
\begin{equation}
{E_{\rm ch} \over L} = -[\gamma_{\theta} k_0 +
\gamma_n k_0 (k_0 R)^2 ] \pi R^2 .
\label{chenergy}
\end{equation}
Equations (\ref{bulk}) and (\ref{chenergy}) can be used to determine the
twist wavenumber $k_0$ for a cylindrical rod of radius $R$.
\par
\begin{figure}
\centerline{\epsfbox{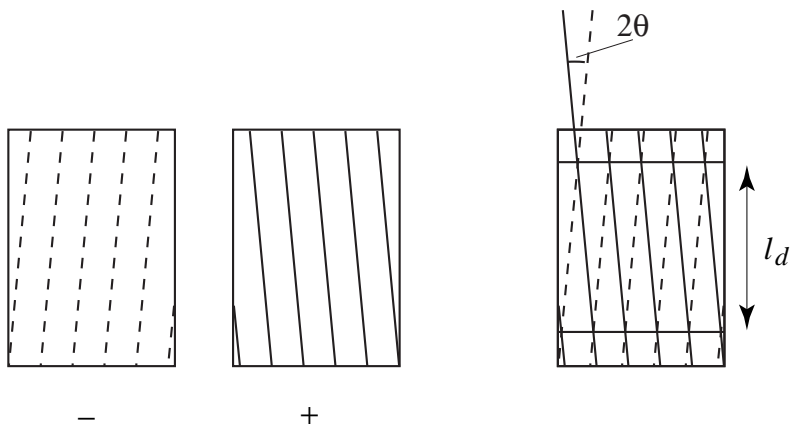}}
\caption{Creation of a twist grain boundary at the interface between two
cells. All cells twist in the same direction.  The columns in back (-) and in
from (+) of a given interface twist in opposite directions.  The interface is
thus a twist grain boundary consisting of screw dislocations
perpendicular to the $z$ axis separated by a distance $l_d =   d/2 \sin
\theta$, where $2 \theta$ is the change in angle of the columnar lattice
across the boundary.}
\label{fig6}
\end{figure}

Our goal is to determine the lowest energy lattice of rods for given values
of the coupling constants.  We begin by distorting our cylindrical rods into
hexagonal rods so that they can pack closely in an hexagonal array (Fig.\
\ref{fig5}).  Each unit cell of the lattice is now a hexagonal cylinder of
twisted columns.  This
distortion will cost an energy for each rod proportional to its shear
modulus times its volume $\pi R^2 L$: $E_{\rm dist.} = \alpha \mu \pi R^2 L$,
where $\alpha $ is a constant of order unity.
The columns at the outer edge of the hexagonal unit
cells make an angle of $\theta = \tan^{-1}
k_0 R \approx k_0 R$ relative to the vertical $z$ axis.
Columns on opposite sides of
a boundary between adjacent hexagonal unit cells, however, make opposite angles
relative to this axis, so the boundary is in fact a twist grain boundary
across which the columnar lattice undergoes a change in angle of $2 \theta$
as shown in Fig.\ \ref{fig6}.
There is a positive energy cost to a grain boundary coming from the cost of
creating the dislocations in the boundary.  There is also an energy gain in
chiral systems arising from the $-\gamma_n \nv \cdot (\gradv \times \nv)$
term in $f_{\rm ch}$.  Consider a boundary in the $xz$ plane with dimensions
$L_x \times L_z$. Since $\int dy \nv \cdot \gradv \times \nv = L_x L_z2
\theta$,
where the integral is across the grain boundary, there is a chiral energy
gain of $-\gamma_n L_x L_z 2 \theta \approx -\gamma_n L_x L_z 2 k_0 R$.
The dislocation energy is $E_{\rm disc} = N_d L_x
\epsilon$, where $\epsilon$ is the energy per unit length of a
dislocation, and $N_d = L_z/l_d$ is the number of dislocations with $l_d$
the distance between dislocations.  There is a simple geometric relation
between $l_d$ and $\theta$: $\theta =  \sin^{-1} d/2 l_d \approx  d/2 l_d$,
where $d$ is the spacing between columns in the columnar lattice.  Thus the
energy of a grain boundary for small $k_0R$ is $E_{\rm GB} = L_x L_z (\epsilon
d^{-1} - 2\gamma_n)2 k_0 R$. The length $L_x$ is proportional to $R$:
$L_x= a\pi R/3$, where $a$ is a constant of order unity.
The first term in $f_{\rm ch}$
contributes an energy $-\pi R^2 L_z \gamma_{\theta} k_0$ per column.
Finally, there is an energy per unit length  $\tau/2$ for each vertex of the
hexagonal lattice where grain boundaries meet. Combining all of these
energies and keeping only the dominant terms, we find
\begin{equation}
f = {E \over V} = {\overline \mu} (k_0 R)^4 - \gamma k_0 + {\tau \over R^2} +
\alpha \mu ,
\end{equation}
where $\gamma = \gamma_{\theta} + (4 a/3)( \gamma_n - \epsilon d^{-1})$.
Minimization of this energy over $k_0$ for $\gamma >0$ yields
\begin{equation}
k_0 =\left( {\gamma \over 4 {\overline\mu} R^4} \right)^{1/3} \sim
R^{-4/3} .
\end{equation}
Thus $k_0 R \sim R^{-1/3}$ is much less than one when $\gamma$ is small
and $R$ is large.  At this optimal value for $k_0$,
\begin{equation}
f = - {3 \over 4} \gamma \left({\gamma \over 4 {\overline\mu}}
\right)^{1/3} R^{-4/3} + {\tau \over R^2} + \alpha \mu .
\end{equation}
Minimization over $R$ yields
\begin{equation}
R = {(8 {\overline \mu} \tau )^{3/2}\over \gamma^2}
\end{equation}
and
\begin{equation}
f = - {1 \over 64} {\gamma^4 \over {\overline \mu} \tau^2} + \alpha \mu .
\end{equation}
Thus, for sufficiently strong chiral coupling $\gamma$, this free energy will
be lower than that of a uniform columnar phase with no twist, and it is
possible that there is some range of parameters for which this phase has a
lower energy than either the moir\'{e} or the soliton phase.
\par
ABH was supported by NSF Grant Number 95-20175.  RDK and TCL were
supported by NSF Grant Number DMR94-23114 and by the Materials Science and
Engineering Center Program of NSF under award number DMR96-32598.

\end{document}